\documentclass[apl%
,aip%
,amssymb%
,amsmath%
,showpacs%
,superscriptaddress%
,reprint%
]{revtex4-1}
\usepackage{graphicx}

\begin{document}
\title{Dipolar ordering of random two-dimensional spin ensemble
}
\author{Andrey V. Panov}
\email{panov@iacp.dvo.ru}
\affiliation{Institute of Automation and Control Processes, Far East Branch of Russian Academy of Sciences, 5, Radio st., Vladivostok, 690041, Russia}

\begin{abstract}
We theoretically study the randomly positioned two-dimensional system of interacting magnetic dipoles representing the monolayer arrays of single-domain particles.
It is showed the onset of the dipolar in-plane superferromagnetic ordering of Ising spins at the surface concentrations of nanoparticles above 0.24. The critical concentration of particles with random easy axis arrangement at zero temperature is 0.65. It is demonstrated that the ensemble with Ising arrangement of spins is ordered at high temperatures close to a particle Curie point.

\noindent Journal reference: Appl. Phys. Lett. 100, 052406 (2012); doi:10.1063/1.3681788
\end{abstract}
\maketitle

Dipolar ferromagnetism, due to interactions between the magnetic moments of rare-earth atoms in crystal lattices, was first observed at very low temperatures \cite{Roser90}. Recently, ordering associated with dipole-dipole coupling between monodomain particles randomly placed on a surface was discovered at high temperatures \cite{Yamamoto08,Fang10,Yamamoto11}. The nanoparticle in such systems in view of its small size acts as a single spin.

Zhang and Widom \cite{Zhang95} applying generalized mean-field theory revealed that the three-dimensional random system of spins within infinitely prolated sample can have dipolar ordering. However, the dipolar ferromagnetism in a moderately elongated spheroid is impossible \cite{Belokon01:eng}. From geometrical considerations, one can expect in-plane dipolar ferromagnetism in the two-dimensional systems. Using a point source model which is only valid at the vanishing spin concentrations, Meilikhov and Farzetdinova \cite{Meilikhov04} showed the absence of the dipolar ordering in the random two-dimensional system of Ising dipoles. Nevertheless, it was demonstrated in Ref.~\cite{Meilikhov04} employing Monte Carlo modeling a percolative transition to the dipolar ferromagnetism in square lattice. Scheinfein et al. \cite{Scheinfein96} studied by the mean-field theory and Monte Carlo simulations the ordering of random Fe islands on the surface and concluded that dipole-dipole interactions are insufficient to explain room temperature magnetism observed.

In this work, by means of generalized mean-field theory, we show the possibility of in-plane dipolar ordering in the two-dimensional sufficiently closely packed system of the spins.

Let us consider the random two-dimensional monolayer system of identical single-domain particles. In the first approximation, we can suppose the nanoparticles as magnetic dipoles which interact with each other spin. At surface concentrations of cylindrical nanoparticles $c=N s/S$ ($s$ is the area of the particle, $S$ is the area of the sample, $N$ is the number of the particles) as high as $0.6$ projection $H$ of the net random field of all the dipoles onto selected direction satisfies the Gaussian distribution \cite{surfdip}:
\begin{equation}
 W_G(H) = \frac{\exp(-(H+H_0)^2/2\sigma^2)}{\sqrt{2\pi}\sigma}
\label{gaussianh}
\end{equation}
with
\begin{equation}
 H_0=-\frac{c}{s}\int \xi \tau(\mathbf{m})d \mathbf{m}\, d\mathbf{r},\:
 \sigma^2=\frac{c}{s}\int \xi^2 \tau(\mathbf{m})d \mathbf{m}\, d\mathbf{r},
 \label{defsigma}
\end{equation} 
where $\mathbf{r}$ is the two-dimensional radius-vector, $\tau \left( \mathbf{m} \right)$ is the distribution function over single-domain particle magnetic moments $\mathbf{m}$, whose absolute values are supposed to be equal to $M_s v$ for all the particles, $M_s$ is the saturation magnetization, $v=s h$ is the volume and $h$ is the particle height, $s=\pi r_0^2$, $r_0$ is the particle radius,
\begin{equation}
 \xi=\frac{3(\mathbf{r}\cdot\mathbf{m})\mathbf{r}-\mathbf{m}r^2}{r^5}.
\end{equation}
Integration over radius $r$ in Eq.~%
\ref{defsigma} begins from $2r_0$, which is the minimal possible distance between two particle centers.

We first consider the in-plane Ising model. The magnetic moments of the particles may have two opposite directions lying in the plane of the monolayer, $N\alpha$ dipoles are oriented along a selected direction and $N\beta$ have opposite alignment, $\alpha+\beta=1$, $m=\alpha-\beta$ is an order parameter. For Ising model, the distribution over magnetic moments is given by
\begin{equation}
\tau(\gamma)={\alpha\delta(\gamma)+\beta\delta(\gamma-\pi)},
\end{equation}
where $\gamma$ is the angle between $\mathbf{m}$ and the selected direction.

The parameters of the Gaussian distribution are readily obtained:
\begin{equation}
 H_{0\parallel}=-\frac{M_s v \pi c m}{2r_0 s}=\frac{h_0 c m}{s},\quad
 \sigma_\parallel^2=\frac{11\pi c M_s^2 v^2}{16s(2r_0)^4}=\mu_2 \frac{c}{s}.
\end{equation} 
The order parameter can be found from the next equation,
\begin{equation}
 m=\int_{-\infty}^\infty\tanh(M_s v H/k_B T)W(H)\,dH,
 \label{magneqising}
\end{equation} 
where $T$ is the temperature and $k_B$ is the Boltzmann constant.

The most interesting critical parameter for such a system is critical concentration above which the ensemble has superferromagnetic ordering. Assuming that $m$ is small in Eq.~\ref{magneqising}, we obtain the equation for critical concentration $c_c$:
\begin{equation}
 -\frac{2 h_0}{\mu_2^{3/2}}\sqrt{\frac s {\pi c_c}}\int_{-\infty}^\infty\tanh\frac{M_s v H}{k_B T}\exp\left[ -\frac{H^2 s}{\mu_2 c_c}\right] H \,dH=1.
 \label{tceq}
\end{equation} 
This equation can be also used for the calculation of the critical temperature at fixed concentration.
At zero temperature, $\tanh(M_s v H/k_B T)$ in Eqs.~\ref{magneqising} and \ref{tceq} may be replaced by sign function $\mathop{\mathrm{sgn}}H$. In this case, it is possible to derive the analytic relation for the critical concentration:
\begin{equation}
 c_0=\frac{\pi \mu_2 s}{4 h_0^2}.
 \label{ccrist0}
\end{equation}
Numerical value for $c_0$ does not depend on material and is equal to ${11}\pi/{128}\approx 0.27$. As we can see concentration for exploiting the Gaussian distribution $c\approx 0.6$ exceeds this value. In order to refine calculations, we can represent the negative logarithm of a characteristic function as a series expansion  and obtain the distribution over the dipolar field after inverse Fourier transform \cite{surfdip},
\begin{multline}
W(H) =\frac{1}{2\pi}\int_{-\infty}^\infty\exp\left\lbrace-\frac{c}{s}\left[ \mu_1 i \rho+\frac{1}{2}\sigma^2\rho^2+\frac{i}{3!}\mu_3\rho^3-\right.\right.\\
\left. \left. \frac{1}{4!}\mu_4\rho^4-\frac{i}{5!}\mu_5\rho^5+\ldots+\frac{1}{10!}\mu_{10}\rho^{10}+\ldots\right]-i\rho H  \right\rbrace \,d\rho,
\label{wseriesh}
\end{multline}
where moments $\mu_n$ of function $\tau(\mathbf{m})$ are defined as follows:
\begin{equation}
\mu_n=\int \xi^n \tau(\mathbf{m})d \mathbf{m}\, d\mathbf{r}.
\end{equation}

In our computations, we limited the expansion by the tenth degree of $\rho$. It should be emphasized that  odd-numbered moments $\mu_n$ depend linearly on $m$. Substituting Eq.~(\ref{wseriesh}) into Eq.~(\ref{magneqising}) and assuming that $m$ is small near the critical point, we deduce following equation to calculate $c_c$:
\begin{multline}
\frac{c_c}{2\pi s}\int_{-\infty}^\infty dH\tanh\frac{M_s v H}{k_B T}\int_{-\infty}^{\infty}d\rho\,\sin(\rho H)\rho\\
{}\times\left[{\mu_1}+ \frac{1}{3!}\mu_3 \rho^2- \frac{1}{5!}\mu_5 \rho^4+ \frac{1}{7!}\mu_7 \rho^6-\frac{1}{9!}\mu_9 \rho^8\right]\\
 {}\times \exp\left\lbrace - \rho^2 \left[ \frac{\mu_2}{2}-\frac{1}{4!} \mu_4 \rho^2+ \frac{1}{6!}\mu_6 \rho^4\right. \right. \\
 \left. \left. {}-\frac{1}{8!}\mu_8 \rho^6+\frac{1}{10!}\mu_{10} \rho^8\right] \frac{c_c}{s}\right\rbrace =1.
 \label{tceqexpasion}
\end{multline}
For the in-plane Ising ensemble of spins the moments are:
\begin{align*}
\mu_3&=\frac{29\pi M_s^3 v^3 m}{56 (2 r_0)^7 }, &&&
\mu_4&=\frac{467\pi M_s^4 v^4}{640 (2 r_0)^{10} }, \\
\mu_5&=\frac{1583\pi M_s^5 v^5 m}{1664 (2 r_0)^{13} }, &&&
\mu_6&=\frac{12031\pi M_s^6 v^6}{8192 (2 r_0)^{16} }, \\
\mu_7&=\frac{44053\pi M_s^7 v^7 m}{19456 (2 r_0)^{19} }, &&&
\mu_8&=\frac{121321\pi M_s^8 v^8}{32768 (2 r_0)^{22} }, \\
\mu_9&=\frac{5026907\pi M_s^9 v^9 m}{819200 (2 r_0)^{25} }, &&&
\mu_{10}&=\frac{38319493\pi M_s^{10} v^{10}}{3670016 (2 r_0)^{28} }.
\end{align*}

Eq.~(\ref{tceqexpasion}) can be solved numerically. Using the parameters of the magnetic nanoparticles observed experimentally \cite{Yamamoto08,Yamamoto11}, we calculated that for Co particles $c_0\approx 0.2376$ and for magnetite $c_0\approx 0.2374$. These values differ little from $c_0$ obtained using the Gaussian distribution.

\begin{figure}[t]
{\centering\includegraphics[width=7.5cm,keepaspectratio=true]{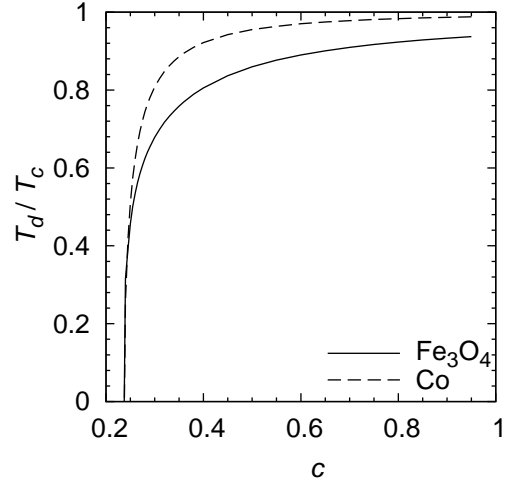} \par}
\caption{Phase diagrams for in-plane Ising dipoles. The areas under the curves correspond to the ordered state, $T_c$ is the Curie temperature of nanoparticle material, $T_d$ is the critical temperature of dipolar ordering.\label{fig:surfdipisingexpand}}
\end{figure}
The phase  diagrams showing conditions for the dipolar alignment of in-plane Ising spins are illustrated in Fig.~\ref{fig:surfdipisingexpand}. The curves were calculated by virtue of solving Eq.~(\ref{tceqexpasion}). For the numerical simulation of cobalt particles, we utilized the dependence of $M_s$ on temperature proposed in Ref.~\cite{Kuzmin05} with Curie temperature $T_c=1385$~K. The magnetite nanoparticles was modeled using expression for $M_s(T)$ based on the Weiss theory:
\[
M_s = 480 \sqrt{\frac{848 - T}{548}}.
\]
It is evident from Fig.~\ref{fig:surfdipisingexpand} that temperature $T_d$ of the Ising ensemble dipolar ordering is mostly constrained by the Curie point of the particles.

Further, we consider the two-dimensional ensemble of the single-domain particles with random orientation of their easy axes. We suppose that the selected direction ($z$-axis) again lies in the plane and the particle magnetic moments may acquire two opposite directions. Now, $N\alpha$ is the number of spins pointing into half-space with positive $z$ and $N\beta$ dipoles point into another half-space, $m=\alpha-\beta$. In this case, the distribution over magnetic moments is
\begin{equation}
\tau(\gamma,\psi)=\frac{\alpha\sin(\gamma)-\beta\cos(\gamma)}{2\pi},
\end{equation}
where $\psi$ is the azimuthal angle of $\mathbf{m}$, $\gamma$ varies within range $[0,\pi/2]$
and the equation for determining magnetization becomes
\begin{equation}
 m=\int_0^{\pi/2}\sin\theta\,d\theta\int_{-\infty}^\infty\tanh\frac{M_s v H\cos\theta}{k_B T}W(H)\,dH
 \label{magneqrandomdef}
\end{equation} 
After the integration over $\theta$ we arrive at following equation:
\begin{equation}
 m=\int_{-\infty}^\infty\ln\left[\cosh\frac{M_s v H}{k_B T}\right]\frac{k_B T}{M_s v H}W(H)\,dH
 \label{magneqrandom}
\end{equation} 
The equation for the critical concentration calculation with Gaussian as $W(H)$ turns to
\begin{equation}
 -\frac{2 h_0 k_B T}{\mu_2^{3/2} M_s v}\sqrt{\frac s {\pi c_c}}\int_{-\infty}^\infty\ln\left[\cosh\frac{M_s v H}{k_B T}\right]\exp\left[ -\frac{H^2 s}{\mu_2 c_c}\right] \,dH=1.
 \label{tceqrandom}
\end{equation} 
The parameters of the Gaussian distribution of the system of the particles with the random easy axis orientation are
\begin{equation}
 H_{0}=-\frac{M_s v \pi c m}{4r_0 s}=\frac{h_0 c m}{s},\quad
 \sigma^2=\mu_2 \frac{c}{s}=\frac{5\pi c M_s^2 v^2}{12s(2r_0)^4}.
\end{equation} 

\begin{figure}[t]
{\centering\includegraphics[width=7.5cm,keepaspectratio=true]{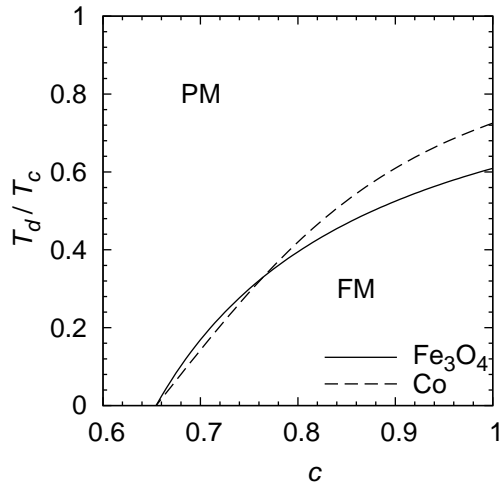} \par}
\caption{Phase diagrams for dipoles with random orientation of axes. The areas under the curves correspond to the ordered state, $T_c$ is the Curie temperature of nanoparticle material, $T_d$ is the critical temperature of dipolar ordering.\label{fig:surfdiprandom}}
\end{figure}

At zero temperature we come again to formula (\ref{ccrist0}), which gives now $c_0={5}\pi/{24}\approx 0.65$. This value 
is appropriate for making use of the Gaussian distribution of the dipolar field. Fig.~\ref{fig:surfdiprandom} shows the phase diagrams calculated with Eq.~\ref{tceqrandom}. We can see from this figure that the critical temperatures are below than in the case of parallel Ising dipoles. This can be attributable to the chaotic arrangement of spins and, hence, their fields preventing ordering.

\begin{figure}[t]
{\centering\includegraphics[width=7.5cm,keepaspectratio=true]{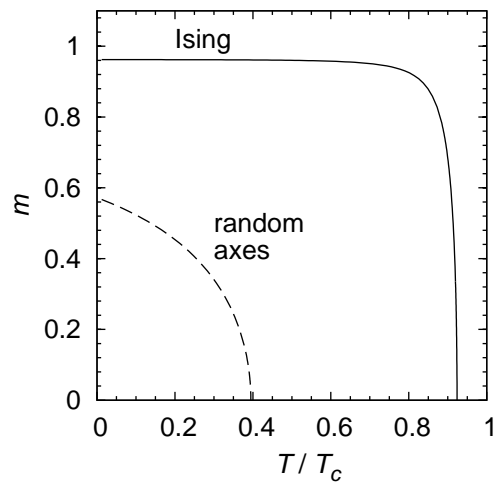} \par}
\caption{Dependence of the order parameter of magnetite nanoparticle two-dimensional ensemble on temperature for $c=0.8$.\label{fig:msurfdip}}
\end{figure}

Fig.~\ref{fig:msurfdip} depicts the order parameters of magnetite particle system as a function of temperature calculated using Eqs.~(\ref{magneqising}) and (\ref{magneqrandom}). Obviously, the degree of magnetic ordering and the critical temperature of random axis spin ensemble lie much below than ones of the Ising spin system. The experimentally observed distribution of magnetite nanoparticle magnetic moments was between the random axis and Ising arrangements \cite{Yamamoto11}. Hence, our calculations do not contradict these experimental data.

Ferromagnetically ordered two-dimensional nanostructures examined in Refs.~\cite{Scheinfein96,Fang10} have surface concentrations $c\approx 0.5$ which are below $c_0$ of the random axis model. Scheinfein et al. \cite{Scheinfein96} pointed out that exchange is responsible for the order in their experiments. It seems that two-dimensional Co nanodot assemblies fabricated by Fang et al.~\cite{Fang10} has anisotropy due to growth process. Monte Carlo algorithm \cite{Du10} used in Ref.~\cite{Fang10} also accounts for the anisotropy. Thus, these data do not quite correspond to the random axis model which has the higher critical concentration for dipolar ordering onset.

It worth mentioning that in the case of spins oriented perpendicularly to the plane $H_{0\perp}>0$ (Ref.~\cite{surfdip}) so that only antiferromagnetic ordering may exist. Obviously, this fact arises from the demagnetizing field of the planar dipole system which has maximum in the transverse direction. It should be also emphasized that the in-plane order in the two-dimensional system has local character limited by a certain size \cite{Imry75}. The experimental results show splitting into domains \citep{Yamamoto11}.

In summary, we theoretically demonstrated the possibility of dipolar ferromagnetism in the two-dimensional random system of spins. We found the critical concentrations of finite dipoles at which this order sets in. It follows from the results of the modeling that the system with in-plane anisotropy has dipolar order at high temperatures. 

The author is grateful to Dmitri Feldman for useful discussion.

\bibliography{surfmagn}

\end{document}